\documentclass{aa}
\usepackage{graphicx}
\begin{document}

%
   \title{Physical-chemical modeling of the low-mass protostar 
     IRAS 16293-2422 
	  } 


   \author{S. D. Doty \inst{1}
           \and
           F. L. Sch\"oier \inst{2,3}
	   \and
	   E. F. van Dishoeck \inst{3}
          }

   \offprints{S. Doty}

   \institute{Department of Physics and Astronomy,
              Denison University, Granville, OH  43023
          \and
	      Stockholm Observatory, 
	      AlbaNova, SE-10691, Stockholm,
	      Sweden
          \and
              Leiden Observatory, P.O. Box 9513,
              2300 RA Leiden, The Netherlands
             }

   \date{Received 9 October 2003; Accepted 29 January 2004}

   \titlerunning{Chemistry in IRAS 16293-2422}

   \abstract{ We present detailed gas-phase chemical models for the
   envelope of the low-mass star-forming region IRAS 16293-2422.  By
   considering both time- and space-dependent chemistry, these models
   are used to study both the physical structure proposed by Sch\"oier
   et al.  (\cite{schoieretal2002}), as well as the chemical
   evolution of this region.  A new feature of our study is the use of
   a detailed, self-consistent radiative transfer model to translate
   the model abundances into line strengths and compare them directly with
   observations of a total of 76 transitions for 18 chemical species,
   and their isotopes.  The model can reproduce many of the line
   strengths observed within 50\%.
   The best fit is for times
   in the range of $3\times 10^3 - 3\times 10^4$ yrs and requires
   only minor modifications to our model for the high-mass star-forming
   region AFGL 2591.  The ionization rate for the source may be higher than
   previously expected -- either due to an enhanced cosmic-ray
   ionization rate, or, more probably, to the presence of X-ray induced
   ionization from the center.  A significant fraction of the CO is found 
   to desorb in the temperature range of 15--40~K; below this temperature 
   $\sim$90\% or more of the CO is frozen out. The inability of the 
   model to explain the HCS$^{+}$, C$_{2}$H, and OCS 
   abundances suggests the importance of
   further laboratory studies of basic reaction rates.  Finally,
   predictions of the abundances and spatial distributions
   of other species which could be observed by future
   facilities (e.g. Herschel-HIFI, SOFIA, millimeter arrays) are 
   provided.  
  \keywords{Stars:
   formation -- Stars: individual: IRAS 16293-2422 -- ISM: molecules }
   }
   
   \maketitle

%

\section{Introduction}
The distribution and composition of dust and gas around isolated
low-mass young stellar objects (YSOs) is receiving increased attention
both observationally and theoretically.  While the general process of
low mass star formation is relatively well understood (see e.g. Shu et
al. \cite{shu1993}; Evans \cite{evans1999}, and others), many details
on the chemical and physical structure at different stages of
evolution remain uncertain.  In particular, the warm and dense gas in
the very interior of these star-forming regions provides a rich
chemical environment with which to probe their structure, properties,
and evolution.  Beyond their own intrinsic interest, these regions may
provide a link to the so-called hot cores observed toward many high-mass 
star-forming regions (e.g., Walmsley \& Schilke
\cite{walmsleyetal1993}).

\begin{figure*}
      \resizebox{\hsize}{!}
      {\includegraphics[angle=-90]{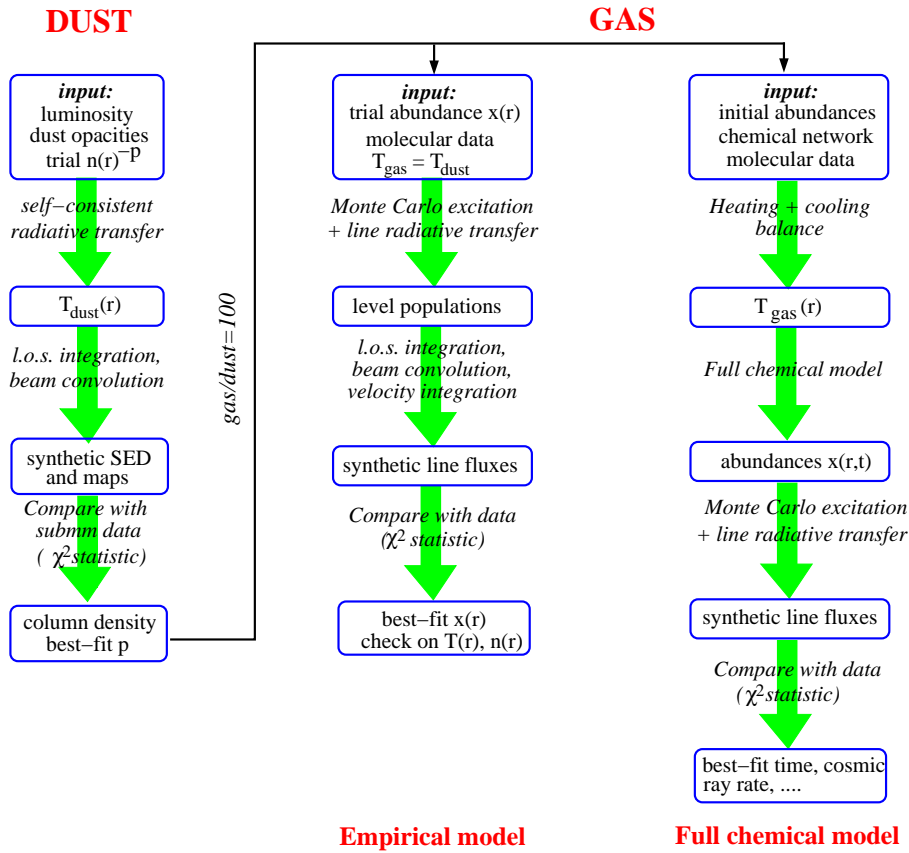}}
      \caption[]{An overview of methods used for constraining the
                 physical and chemical structure of YSO envelopes.
		 In general, the steps include the determination
		 of the physical structure through dust modeling, 
		 calculation of gas temperature, and adoption of
		 a chemical model.  This combination produces 
		 observables (column densities, line fluxes, etc.)
		 to compare with observations.  The source parameters
		 are determined by adjusting them until a best fit
		 is obtained.  Two important points of divergence
		 include the use of self-consistent vs. 
		 approximate radiative transfer, and the use of
		 a time-dependent chemical network 
		 (`Full chemical model') vs. simple
		 trial abundances (`Empirical model') 
		 (figure adapted from 
		 van Dishoeck \& van der Tak \cite{vDandvdTreview2000},
		 and van Dishoeck \cite{vdreview2003}).
              }
         \label{modelingflowchart}
   \end{figure*}

Rapid advances in both observational and modeling capabilities allow
much more quantitative studies of the chemistry in YSO envelopes than
was possible even a few years ago.  Several different steps can be
distinguished (see Fig. \ref{modelingflowchart}). Thanks to the advent
of large-format bolometer arrays, most studies nowadays start with an
analysis of the spatial distribution of the submillimeter continuum
emission from dust and its spectral energy distribution (SED) (e.g.,
Shirley et al.\ \cite{shirleyetal2000}; J{\o}rgensen et al.\
\cite{joergensenetal2002}; Sch\"oier et al.\ \cite{schoieretal2002}). Through
continuum radiative transfer calculations, both the density profile
$n\propto r^{-p}$ as a function of radius $r$ and the dust temperature
structure $T_{\rm dust}(r)$ can be determined self-consistently. For
the gas, two approaches can subsequently be taken. In atmospheric
chemistry, these two cases are commonly known as the `forward' and
`backward' or `retrieval' methods. In the `empirical model'
(`retrieval'), $T_{\rm gas}$ is taken to be equal to $T_{\rm dust}$
and the excitation, radiative transfer and fluxes of the various
molecular lines are calculated for an assumed abundance profile
$x(r)=n({\rm X})(r)/n$(H$_2$)($r$). This trial abundance profile is
then varied until the best agreement with observations is obtained. In
practice, only two types of abundance profiles are considered: a
constant abundance throughout the envelope or a `jump' profile in
which the abundance is increased by a large factor in the inner warm
region due to ice evaporation.  Such models have successfully been
applied to both high- (e.g., van der Tak et al.\
\cite{vdt2000}) and low-mass YSOs (e.g., Ceccarelli et al.\
\cite{ceccarellietal2000b}, Sch\"oier et al.\ \cite{schoieretal2002}),
and work best for molecules for which a large set of lines originating
from levels with a range in energy has been observed.  This method
provides abundances for comparison with chemical models, but does not
test the chemical networks directly.
%
%
%

The second, {\it ab initio} or `forward' approach is the `full
chemical model', in which only the density structure derived from the
dust is adopted as a starting point. Given initial abundances and a
detailed chemical network, the abundances of various molecules can be
solved as functions of position and time and the gas temperature can
be calculated explicitly by solving the full thermal balance of the
gas.  The physical structure $n(r)$ and $T(r)$ can either be taken to
be constant with time or to vary according to some (dynamical)
prescription.
Such time- and space-dependent chemical models have been applied to
low-mass YSOs by Ceccarelli, Hollenbach, \& Tielens (\cite{cht96},
hereafter CHT96) and Rodgers \& Charnley (\cite{rodgerscharnley2003}), and
to specific high-mass sources by Millar, MacDonald \& Gibb
(\cite{millaretal1997}; G34.3+0.15) and Doty et al.\
(\cite{dotyetal2002}; AFGL 2591). The main quantities to be determined
are the best-fit time (or, in a dynamical model, mass-accretion rate)
and other parameters which enter the chemical models, such as the
cosmic ray ionization rate. The chemical models themselves can be
tested by comparing the abundance profiles $x(r,t_{\rm fit})$ at the
best-fit time with those derived through the empirical method. In this
way, they also provide a guideline for more complicated abundance
profiles to adopt in the empirical method.

In this paper, we describe a detailed `full chemical model' of the
best studied low-mass YSO, IRAS 16293-2422. A novel feature is the
addition of a full Monte Carlo radiative transfer calculation of the
resulting line fluxes for direct comparison with observations (see
bottom-right part of Fig. \ref{modelingflowchart}).  Such models
provide the most complete test of our understanding of the physical
and chemical structure of YSO envelopes.

IRAS 16293-2422 is a nearby ($\sim 160$ pc, Whittet
\cite{whittet1974}) low mass, low luminosity (27 L$_{\odot}$)
protostellar object located within the $\rho$ Ophiuchus molcular
cloud.  It has an exceptionally rich and well-studied spectrum (e.g.,
Blake et al. \cite{blakeetal1994}; van Dishoeck et
al. \cite{vandishoecketal1995}; Cecarelli et
al. \cite{ceccarellietal2000a}, \cite{ceccarellietal2000b}; Sch\"oier
et al. \cite{schoieretal2002}), and is therefore considered the
prototypical low-mass source for chemical studies, much like Orion is
for high-mass objects.  Ceccarelli et al.\ (2000a,b) used the physical
structure based on CHT96 combined with a restricted chemical network
to analyze data of H$_2$CO, H$_2$O, and SiO in a `full chemical
model', and found strong evidence for large abundance enhancements of
these species in the innermost part ($\leq$150 AU) of the
envelope. The evaporated species may subsequently drive a complex `hot
core' chemistry leading to the even more complex organic molecules
which have recently been detected in IRAS16293-2422 (Cazaux et
al. \cite{cazauxetal2003}).


In a later analysis, Sch\"oier et al. (\cite{schoieretal2002})
combined dust/SED modeling of the physical structure, multiple line
observations covering a range of excitation conditions, and a detailed
radiative transfer analysis in an `empirical model' to infer the
structure of IRAS 16293-2422.  This work supported the general conclusion
of a `hot core' where the abundances of key molecules are enhanced by
several orders of magnitude due to evaporation of ices.  The model
employed only uniform and `jump' abundances, however, which may not be
representative of the detailed time- and space-dependent chemistry.

Recently, Doty et al. (\cite{dotyetal2002}) described such a time- and
space-dependent physical/chemical model for static YSO envelopes including
the hot core chemistry.  By combining the model results with
observations of many species of one particular high-mass YSO,
AFGL~2591, it has been shown that it may be possible to not only
confirm the gross source structure, but also constrain source
properties such as age, ionization rate, and role of grains in
determining the chemical structure (see also Boonman et
al. \cite{boonmanetal2003}).  Here the `full chemical model' of
Fig. \ref{modelingflowchart} was adopted, but the self-consistent line
radiative transfer was performed for only a subset of the species.

In this paper, we report on the application of the physical/chemical
model of Doty et al. (\cite{dotyetal2002}) to the low-mass YSO IRAS
16293-2422.  These results are combined with a self-consistent
radiative transfer model, and applied to the full multi-species,
multi-transition dataset of Sch\"oier et
al. (\cite{schoieretal2002}). By comparison with the case of
AFGL~2591, we can also directly determine the differences in derived
model parameters for a low- and a high-mass YSO (van Dishoeck 2003).
The models and observations are briefly described in Section 2.  The
observations are then used with the models to constrain the source
properties in Section 3.  Finally, we summarize the results and
conclude in Section 4.

\section{Existing observations and models}

\subsection{Observations}

IRAS 16293-2422 has been well-observed both in the continuum and
in various submillimeter molecular lines.  While no new
observations are presented in this paper, we briefly note
and discuss the observational data as they provide the constraints
placed on the model.

The SED of IRAS 16293-2422 in the range $60-2900$ ${\mu}$m is
presented by Sch\"oier et al. (\cite{schoieretal2002}).  
High angular resolution JCMT data allowed the
determination of radial brightness distributions at $450$ and
$850$ ${\mu}$m on scales of 9$''$ and 15$''$ (1400 and 2400 AU) respectively.

   \begin{table}
      \caption[]{IRAS 16293-2422 physical structure from dust modeling
         by Sch\"oier et al.\ (\cite{schoieretal2002})}
         \label{modelparameters}
         \begin{tabular}{ll}
            \hline
            Parameter & Value \\ 
            \hline
     Distance, $d$ (pc) & 160 \\
     Luminosity, $L$ ($L_{\odot}$) & 27 \\
     Optical depth at $100$ ${\mu}$m, ${\tau_{100}}$ & 4.5 \\
     Density power law index, $p$ & 1.7 \\
     Inner envelope radius, $r_{i}$ (cm) & $4.8 \times 10^{14}$ \\ 
     Outer envelope radius, $r_{e}$ (cm) & $1.2 \times 10^{17}$ \\
     H$_{2}$ density at 1000 AU, $n_{0}$ (cm$^{-3}$) & $6.7 \times 10^{6}$ \\
     
            \hline
     \end{tabular}\\
   \end{table}

The molecular line data utilized here are taken primarily from
the large surveys of IRAS 16293-2422 by Blake et al. 
(\cite{blakeetal1994})  and van Dishoeck et al. 
(\cite{vandishoecketal1995}).
Additional complementary 
data are taken from the JCMT public archive 
(Sch\"oier et al. \cite{schoieretal2002}).  
These data are supplemented by the H$_{2}$CO lines 
of Loinard et al. (\cite{loinardetal2000}).  The 
data set 
-- a total of 76 transitions for 18 species considered here --
has the advantage that it samples the full radial
range of the envelope, providing probes over a wide range
of physical, thermal, and chemical conditions.  
Only information on the lowest transitions of the 
molecules, which occur at millimeter wavelengths and probe
the very coldest outer parts and surrounding cloud, is lacking.
In general, the calibration uncertainty of each individual line is $\sim$30\%.

\subsection{Model}
Here a brief synopsis of the physical, thermal, chemical, and
radiative transfer models are provided.  For more detailed
information, see Doty et al. (\cite{dotyetal2002}), 
Sch\"oier et al. (\cite{schoieretal2002}), Doty \& Neufeld
(\cite{dn97}), and references there.  

\subsubsection{Physical and thermal structure}
\begin{figure}
      \resizebox{\hsize}{!}{\includegraphics{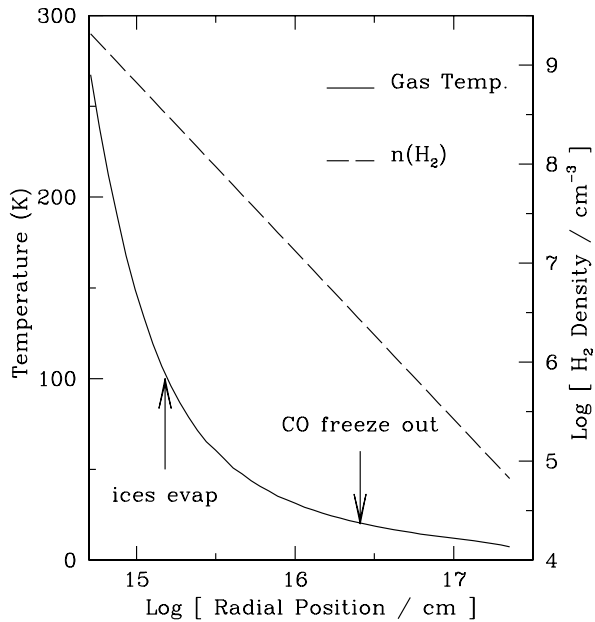}}
      \caption[]{Physical and thermal structure of IRAS 16293-2422.
                 The density and temperature are from the model
		 of Sch\"oier et al. (\cite{schoieretal2002}).  The
		 gas temperature is assumed to follow the dust
		 temperature. 
              }
         \label{tempdens}
   \end{figure}

We adopt a spherically symmetric static model of the extended envelope
of IRAS 16293-2422 surrounding the two protostellar sources in the
center (Looney et al. \cite{looneyetal2000}).  The observational
continuum data were combined and simultaneously modeled by Sch\"oier
et al. (\cite{schoieretal2002}) with the publicly available radiative
transfer code DUSTY (Ivezic et al.\ \cite{ivezicetal1999}).  This
analysis allows the source structure properties (e.g., envelope size,
density power law, continuum optical depth) to be determined to within
approximately $\pm 20$\% (see also Doty \& Palotti
\cite{dotyandpalotti2002}).  The adopted source properties are
presented in Table \ref{modelparameters}, and the density and
temperature structure are reproduced in Fig. \ref{tempdens}

The successful line modeling of Ceccarelli et
al. (\cite{ceccarellietal2000a}) and Sch\"oier et al.\ (2002) in
spherical symmetry down to $\sim 30$ AU suggests that the assumption
of spherical symmetry may be largely justifiable.  Most chemical data
were obtained with a $15''$ beam, which yields a linear size of 2400
AU at the assumed distance of $160$ pc, larger than the 800 AU
separation of the central protostars.  Sch\"oier et
al.\ (\cite{schoieretal2003}) recently used interferometric obesrvations
to determine the structure below 1000 AU, finding that the binary has
cleared out most of the material in the inner part of the envelope,
and that there exist two unresolved central sources with best-fit disk
sizes of $\sim 250$ AU in diameter.  However, they find that the
while the detailed results for the inner envelope leave the
inner ($r < 400$ AU) structure somewhat uncertain, their results have little
effect on the extended envelope ($r > 400$ AU).

%
%
%

A majority of the modeled envelope exists at relatively high densities
and optical depths, leading to a strong thermal coupling between the
gas and dust.  As a result, the gas temperature is assumed to follow
the dust temperature (Ceccarelli et al. \cite{cht96}; Doty \& Neufeld
\cite{dn97}; Ceccarelli et al. \cite{ceccarellietal2000a}).  Test
calculations have shown that this assumption is sufficient for both
the chemistry (Doty et al. \cite{dotyetal2002}) and radiative transfer
through molecular lines (Boonman et al. \cite{boonmanetal2003}).

\subsubsection{Chemistry}
The chemical model is based upon the UMIST gas-phase chemical reaction
network (Millar, Farquhar, \& Willacy \cite{MFW}, hereafter MFW), 
including reactions
to model the hot-core chemistry.  Pseudo time-dependent models of the
chemical evolution over 30 radial grid points were constructed,
providing a time- and space-dependent chemical evolution.  The local
parameters (hydrogen density, temperature, and optical depth) at each
radial point are taken from the physical and thermal structure
calculations above.  For the majority of species, the initial
abundances of the high-mass source AFGL 2591 (Doty et
al. \cite{dotyetal2002}), are utilized as shown in Table
\ref{initialabundances}. The chemistry of deuterated molecules is not
considered.

The effects of freeze out onto and desorption from dust grains are
approximated. Instead of explicit freeze out or desorption with time,
the desorption is taken to be instantaneous at 100 K, where expected
grain mantle species are injected into the gas (Charnley \cite{C97}),
in keeping with the timescales observed in the laboratory (Fraser et
al. \cite{fraserh2opaper}).  Below 100 K, we deplete the gas phase
abundance of many species expected to be in ices (e.g., H$_2$O).  The
only exceptions are CO, which we take to desorb at $T_{\mathrm{CO}}$,
and H$_{2}$CO and CH$_{3}$OH, which we take to desorb or `jump' at
$T_{\mathrm{X}}$.

The effects of photodissociation from the interstellar radiation field
at the outer boundary is included, but is
generally small due to the high optical depth, and the coarseness of
the spatial grid considered.  
Finally, we have considered the effect
of metal depletion by significantly reducing the initial Fe abundance.
We find that metal depletion makes only a small difference, worsening
the fits by only a few percent on average.

   \begin{table}
    \caption[]{Initial abundances at $t$=0 relative to H$_{2}$ for AFGL 2591 }
         \label{initialabundances}
         \begin{tabular}{lll}
            \hline
            Species & Initial Abundance & Ref. \\ 
            \hline
     Initial Abundance ($T>100\mathrm{K}$) & {} & {} \\
     CO & $3.7(-4)$ & a \\
     CO$_{2}$ & $3.0(-5)$ & d \\
     H$_{2}$O & $1.5(-4)$ & d \\
     H$_{2}$S & $1.6(-6)$ & h \\
     N$_{2}$  & $7.0(-5)$ & e \\
     CH$_{4}$ & $1.0(-7)$ & e \\
     C$_{2}$H$_{4}$ & $8.0(-8)$ & e \\
     C$_{2}$H$_{6}$ & $1.0(-8)$ & e \\
     O & $0.0(0)$ & e \\ 
     H$_{2}$CO & $1.2(-7)$ & e \\
     CH$_{3}$OH & $1.0(-6)$ & e \\
     S  & $0.0(0)$ & e \\
     Fe & $2.0(-8)$ & e \\
     {} & {} & {} \\
     Initial Abundance ($T<100\mathrm{K}$) & {} & {} \\
     CO & $3.7(-4)$ & a \\
     CO$_{2}$ & $0.0(0)$ & f \\
     H$_{2}$O & $0.0(0)$ & f \\
     H$_{2}$S & $0.0(0)$ & f \\
     N$_{2}$  & $7.0(-5)$ & e \\
     CH$_{4}$ & $1.0(-7)$ & e \\
     C$_{2}$H$_{4}$ & $8.0(-8)$ & e \\
     C$_{2}$H$_{6}$ & $1.0(-8)$ & e \\
     O & $8.0(-5)$  & g \\
     H$_{2}$CO & $0.0(0)$ & f \\
     CH$_{3}$OH & $0.0(0)$ & f \\
     S  & $6.0(-9)$  & h \\ 
     Fe & $2.0(-8)$ & e \\
            \hline
     \end{tabular}\\
$^{\mathrm{ }}$ { } $a(b)$ means $a \times 10^{b}$\\ 
$^{\mathrm{ }}$ All abundances are gas-phase, and relative to H$_{2}$ \\
$^{\mathrm{a}}$ van der Tak et al. \cite{vdt1999}, 
$^{\mathrm{b}}$ van der Tak et al. \cite{vdt2000}, 
$^{\mathrm{c}}$ van der Tak \& van Dishoeck \cite{vv2000}, 
$^{\mathrm{d}}$ Boonman et al. \cite{boonmanetal2000}, 
$^{\mathrm{e}}$ Charnley \cite{C97}, 
$^{\mathrm{f}}$ assumed frozen-out or absent in cold gas-phase, 
$^{\mathrm{g}}$ taken to be $\sim$ consistent with Meyer, Jura, \& Cardelli 1998
$^{\mathrm{h}}$ Doty et al. \cite{dotyetal2002}
   \end{table}

\subsubsection{Radiative transfer}
The molecular line radiative transfer is accomplished through a
non-LTE, Monte-Carlo model described in Sch\"oier
(\cite{schoier2000}).  This code has been benchmarked to high accuracy
against a wide range of other molecular line radiative transfer models
(van Zadelhoff et al. \cite{rtbench2002}).  In this model, the spatial
molecular abundances $x(r,t)$ are combined with the adopted physical
structure to compute the excitation and resulting line profiles for
all transitions up to $\sim 500$ K in the ground vibrational state of
the observed molecules.  Chemical evolution times from $3 \times
10^{2}$ years to $3 \times 10^{5}$ years are considered, with one dex
spacing.

\section{Results}
In this section the results of our physical/thermal/
chemical modeling of IRAS 16293-2422 are presented, and the comparison of
line strengths predicted from this model to those observed.
As a metric of the comparison, we adopt the mean percentage 
magnitude difference between the predicted and observed line strengths,
given by
\begin{eqnarray}
\Delta =  \frac{1}{N_{\rm{lines}}} 
\sum_{i=1}^{N_{\mathrm{lines}}} 
\left| \frac{F_{\rm{mod},i}-F_{\rm{obs},i}}{F_{\rm{obs},i}} \right|.
\label{percentdiff}
\end{eqnarray}

This form has the advantage that it measures the size of the
difference, without allowing equally balanced high and low
values to cancel out.  Furthermore, the summation over lines
implicitly provides a greater weight to molecules for which 
more data -- and thus more constraints -- exist.  Measures
using the sum over molecules instead of lines show 
qualitatively similar, though often accentuated, results.
We note that we adopt this measure instead of a $\chi^2$ analysis,
as agreement to within a factor of 10 are considered good for
chemical modeling.
This value is enough larger than the statistical 
uncertainty in the line observations so as to invalidate the
statistical meaning of $\chi^2$ measure.

The parameters varied in the models are the cosmic ray
ionization rate $\zeta$, the adopted initial abundances in the inner and outer
regions, and the desorption temperatures of selected species (CO,
H$_2$CO, CH$_3$OH). 
Detailed radial profiles of selected species are
presented in \S 3.7 and 3.8.

\subsection{General Results}
As discussed above, our base model is taken after the high-mass hot
core + envelope of AFGL~2591 by Doty et al.  (\cite{dotyetal2002}).  A
parameter space search, guided by results from previous studies,
suggests a best fit to the observed data of IRAS 16293-2422 with only
minor modifications to the AFGL 2591 initial chemical conditions.

%
   \begin{figure}
      \resizebox{\hsize}{!}{\includegraphics{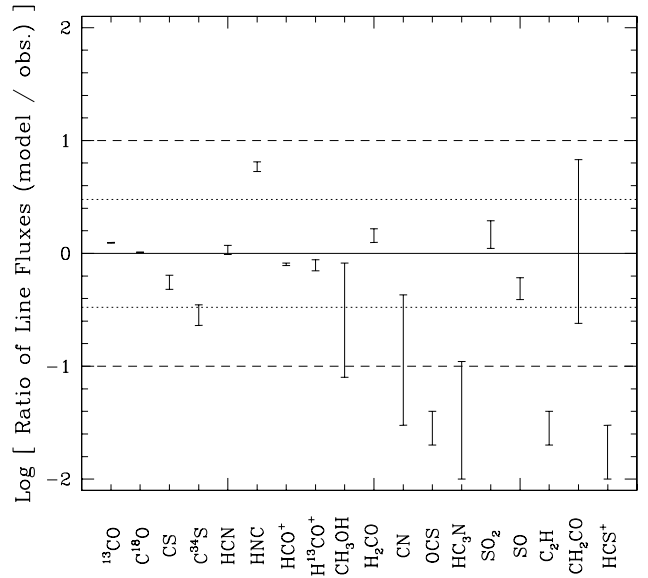}}
      \caption[]{The ratio of the
                 predicted and observed line strengths for
		 molecules observed toward IRAS 16293-2422.
		 The errorbars represent the range of values 
		 between
		 $3 \times 10^{3} < t(\mathrm{yrs}) < 3 \times 10^{4}$.
		 This represents the best fit time range, and
		 is consistent with previous values based upon
		 estimates of the infall rate and central mass.
		 Guidelines at factor of 3 (dotted) and 10 (dashed)
		 ratios are given to indicate good and acceptable fits.
	      }
         \label{bestfitcomparison}
   \end{figure}
%
%

A comparison between the best fit model and observations is shown in
Fig. \ref{bestfitcomparison}.  Here the ratio of 
the predicted to observed line strengths for each molecule observed
is plotted.  We find a best-fit time of $3 \times 10^{3} <
t(\mathrm{yrs}) < 3 \times 10^{4}$, with these times forming the range
in the figure.  These times are consistent with the age inferred by
Sch\"oier et al. (\cite{schoieretal2002}) from fitting a collapse
model to the line profiles, and by using the constant infall rate of
Ceccarelli et al.  (\cite{ceccarellietal2000a}) with their preferred
central mass of $0.8$ M$_{\odot}$.

As can be seen in the figure, the majority of species 
(11 of 18) are fit to within 50 \% of the observations,  
thirteen are fit to within a factor of three, and
15 are fit to within a factor of 10,  
a level usually considered acceptable agreement in chemical modeling
(see e.g., Millar \& Freeman \cite{millarfreeman1984};
Brown \& Charnley \cite{browncharnley1990}; 
Terzieva \& Herbst \cite{terzievaherbst1998}).
An interesting case is the $\sim 20-30\%$ deviation in $^{13}$CO, and 
the small uncertainties on $^{13}$CO, C$^{18}$O, and HCO$^{+}$.
It is possible that the $^{13}$CO discrepancy could be due to a
different $^{12}$C/$^{13}$C ratio than taken here.  It may also
be due to deviation of the real structure from the continuous, spherical
symmetry we have adopted.  In any case, the deviation is no larger than
the expected calibration uncertainty of $\sim 20-30\%$.

In the comparison there exist three species which deviate
by more than an order of magnitude.  The outliers are:
OCS, C$_{2}$H, and HCS$^{+}$, which have individual
deviations of a factor 30, 30, and 100 respectively.
While there is variation, our model tends toward producing
too little emission.  {
The difference between the model and observations is $\Delta=0.51$ 
when these are omitted.  Including them raises $\Delta$ to
$0.87$.
This quantitatively confirms the agreement between the model and
observations at the level of a factor of two for most species.

   \begin{table}
      \caption[]{Differences between best fit model for IRAS 16293-2422
                 and AFGL 2591}
         \label{modeldifferencetable}
         \begin{tabular}{lll}
            \hline
            Parameter & IRAS 16293 & AFGL 2591 \\ 
            \hline
     Initial Abundance ($T>100\mathrm{K}$) & {} & {} \\
     CO & $1.0(-4)$ & $3.7(-4)$ \\
     H$_{2}$S & $1.0(-8)$  & $1.6(-6)$ \\
     H$_{2}$CO & $8.0(-8)$ & $1.2(-7)$ \\
     CH$_{3}$OH & $1.5(-7)$ & $1.0(-6)$ \\
     {} & {} & {} \\
     Initial Abundance ($T<100\mathrm{K}$) & {} & {} \\
     CO ($20 < T(\mathrm{K}) < 100$) & $1.0(-4)$ & $3.7(-4)$ \\
     O & $1.0(-4)$ & $8.0(-5)$ \\
     H$_{2}$CO ($60 < T(\mathrm{K}) < 100$) & $8.0(-8)$ & $0.0$ \\
     CH$_{3}$OH ($60 < T(\mathrm{K}) < 100$) & $1.5(-7)$ & $0.0$ \\
     {} & {} & {} \\
     $T_{\mathrm{des}}(\mathrm{K})$ [$x(T<T_{\mathrm{des}}) \sim 0$] 
        & {} & {} \\
     CO & $20$ & $100$ \\   
     H$_{2}$CO & $60$ & $100$ \\
     CH$_{3}$OH & $60$ & $100$ \\
     {} & {} & {} \\
     CR ionziation rate $[\zeta]$ (s$^{-1}$) & $5.0(-16)$ & $5.6(-17)$ \\
            \hline
     \end{tabular}\\
$^{\mathrm{ }}$ { } $a(b)$ means $a \times 10^{b}$\\ 
$^{\mathrm{ }}$ All abundances are gas-phase, and relative to H$_{2}$ \\
   \end{table}

The differences between the AFGL 2591 model and the best fit model
here are summarized in Table \ref{modeldifferencetable}.  As can be
seen, the differences are generally minor.  They are: (1) a large increase
in the cosmic-ray ionization rate by a factor of $>$10 over the
`standard' value of $\sim 10^{-17}$ s$^{-1}$ and the AFGL 2591 value
of $\sim 5.6 \times 10^{-17}$ s$^{-1}$; (2) depletion of CO at low
temperatures ($\sim 20-40$ K); (3) depletion of H$_{2}$CO and
CH$_{3}$OH at moderate temperature ($< 60$ K); and (4) variations in
the initial abundances of a few other species.  We discuss each of
these differences below separately, as variations from the best fit
model.

\subsection{Effects of cosmic ray ionization rate}

The ionization rate inferred in the modeling, 
$\sim 5 \times 10^{-16}$ s$^{-1}$, is much higher than
the `standard' cosmic ray ionization rate of $\sim 10^{-17}$
s$^{-1}$ (e.g., Roberts \& Herbst \cite{robertsherbst2002};
Black \& Dalgarno \cite{blackdalgarno1977}; 
O'Donnell \& Watson \cite{odonnellwatson1974}).  
To see the dependence of the model results on the
ionization rate, Fig. \ref{figzeta} shows the mean 
difference $\Delta$ between the models and observations upon
varying the ionization rate.  The minimum deviation occurs in the range
$\zeta = 5 \times 10^{-16} - 10^{-15}$ s$^{-1}$.  While the two values
near the minimum are essentially indistinguishable,
the ionization rate required for this fit is 50--100 times
higher than the traditional cosmic-ray ionization rate
used in dark cloud models (Lepp \cite{lepp1992}).

%
   \begin{figure}
      \resizebox{\hsize}{!}{\includegraphics{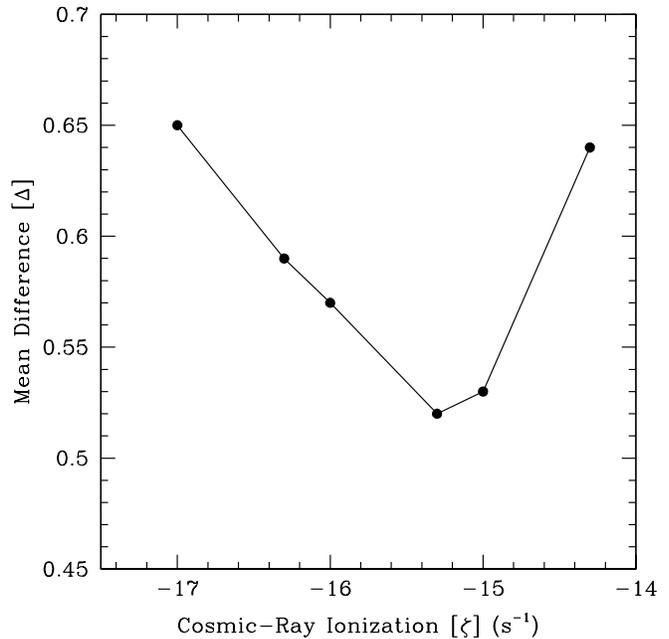}}
      \caption[]{Dependence of quality of fit as measured by
                 the mean difference between predicted and 
		 observed line strengths, as a function of the
		 ionization rate, $\zeta$.
		 Notice the best fit near $\zeta = 5\times10^{-16}$ s$^{-1}$.
		 The two points near the minimum are indistinguishable
		 at the level of uncertainty of the observations, and
		 given the constraints of the model.  In either case,
		 the ionization rate is much larger than standard.
              }
         \label{figzeta}
   \end{figure}
%
%

The best fit range for $\zeta$ is expected to be meaningful, due
to the fact that the mean difference is minimized here.  
The variation in $\Delta$ is damped by the fact that it is an average
across all species.  As a result, a $25\%$ variation in $\Delta$
can correspond to a factor of 2 change in 1/4 of the species,
or a factor of $\sim 8$ change in $\sim 4-5$ transitions.  This
is seen for the case of CO 
in Figures \ref{figco2panel}  and \ref{figcosome} where
a smaller change in $\Delta$ over all species corresponds to 
variation in physical parameters by $\sim 100\times$, and 
variations in line strengths by up to $8 \times$.  As a result, 
we infer that a minimum in $\Delta$ and a variation of $\sim 20\%$
is sufficient to draw conclusions, which implies that the preferred value
of $\zeta = 5 \times 10^{-16} - 10^{-15}$ s$^{-1}$ is meaningfully
different from other values tested.

The ionization rate $\zeta$ is assumed to be uniform throughout the
source.  Species that are most affected by the variation in the
ionization rate are HCO$^{+}$, HCN, SO, and H$_{2}$CO and show
improvements of up to 100\%.  While the HCO$^{+}$ abundance should be
directly related to the ionization rate throughout the envelope, Doty
et al. (\cite{dotyetal2002}) show that the remainder are predominantly
active above $100$ K.  This implies that the ionization rate may be
position dependent, with the most affected species in the interior.

The origin of this enhanced ionization is of physical interest.
Recent models and measurements infer up to 2 orders of magnitude
variation in the cosmic ray ionization rate (McCall et
al. \cite{mccalletal2003}; Liszt \cite{liszt2003}; Doty et
al. \cite{dotyetal2002}; van der Tak \& van Dishoeck \cite{vv2000}).
It is difficult to understand such an extreme variation from source to
source, implying that some other physical mechanism may produce or
contribute to the ionization.  This is especially true as the recent
`high' inferred values for $\zeta$ are for diffuse clouds, while
dense cloud models have historically required much smaller cosmic-ray
ionization rates near $1-3 \times 10^{-17}$ s$^{-1}$.  In the case of
IRAS 16293, we suggest the possibility that the enhanced ionization is
due to X-rays produced by magnetic activity associated with accretion
onto the protostars.  This could both produce the exceptionally high
inferred ionization rate, and preferentially affect the warmer
interior species.  The effects of a central X-ray source will be
presented in a forthcoming paper (Doty et al., in preparation).

\subsection{Effects of CO desorption temperature}

%
   \begin{figure}
      \resizebox{\hsize}{!}{\includegraphics{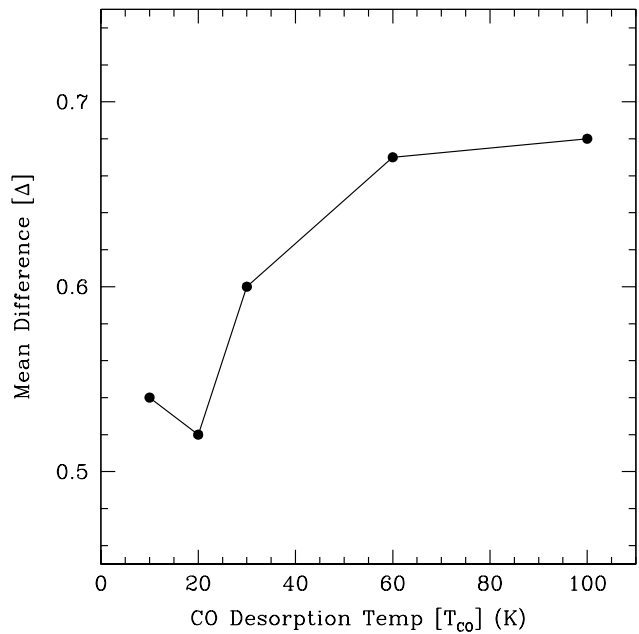}}
      \caption[]{Dependence of quality of fit as measured by the 
                 mean difference between predicted and observed line
		 strengths for the full chemistry / observational set,
		 as a function of the CO desorption temperature 
		  ($T_{\mathrm{CO}}$).  Here, $x(\mathrm{CO}) = 10^{-4}$.
		 Notice the best fit for $T_{\rm CO} \sim 20$ K.
              }
         \label{figtco}
   \end{figure}
%
%

Studies of solid CO on icy dust grains (Collings et
al. \cite{collingsco2003}; Fraser et al. \cite{frasercopaper};
Galloway \& Herbst \cite{gallowayherbst1994}; Sandford \& Allamandola
\cite{sandfordallamandola1993a}; Nair \& Adamson
\cite{nairadamson1970}) suggest that the bulk of the CO evaporates in
a step-wise fashion between 20 and 70~K -- depending upon whether it
is trapped inside or lies on top of the ice -- much lower than the
temperature at which water ice desorbs from grains (Fraser et
al. \cite{fraserh2opaper}).  This is consistent with our best fit
model.  In order to test this, we have varied the CO desorption
temperature from our baseline best-fit model.  The results are shown
in Fig. \ref{figtco}.

Clearly, the best fit requires a CO desorption temperature
near $\sim 20$ K and $< 60$ K.  A lower temperature both yields a worse
fit to the observational data by overproducing the $^{13}$CO and C$^{18}$O 
emission by 25\% and 68\% respectively 
at $T_{\mathrm{CO}}=10$ K, and is inconsistent with
laboratory results.  Much higher temperatures yield significantly
worse fits to the data, underproducing both $^{13}$CO and 
C$^{18}$O line fluxes by a factor of 50 by $T_{\mathrm{CO}}=100$ K.  
The species most affected (aside from
CO itself) are the cyanogens CN, HCN and HNC, and the CO
ion-molecule byproducts HCO$^{+}$, CS, and H$_{2}$CO, all of which 
show variations between 40-300\% .

Physically, a low desorption temperature near 20 K would be an 
indication that a significant fraction of the CO is not 
intermixed with the H$_{2}$O in the grain mantle.
A number of suggestions for differentiation in the ice have been made,
including differentiation in the gas prior to adsorption,
differentiation in the ice due to chemical and physical processing,
and differentiated freeze-out during the cooling time behind a shock
which has liberated the grain mantles (e.g., Schutte \cite{schutte1997}
; Bergin et al.\
\cite{berginetal1999}).  While it is difficult to comment on the first
two scenarios, it is doubtful that shock processing is the main cause
for IRAS 16293-2422.  In particular, the products of shock chemistry
do not dominate the bulk of the envelope, and the small linewidths
observed for many species further suggest that a large fraction of
the volume of the gas is not shocked.  Note that this analysis does
not exclude that some fraction of the CO also evaporates at
higher temperatures.  In fact, J{\o}rgensen et al. 
(\cite{joergensenetal2002})
conclude from their analysis of the CO $3 \rightarrow 2$ and 
$2 \rightarrow 1$ isotopic lines in a sample of class 0 objects that
some CO must still be frozen out at temperatures above 25~K.

%
   \begin{figure}
      \resizebox{\hsize}{!}{\includegraphics{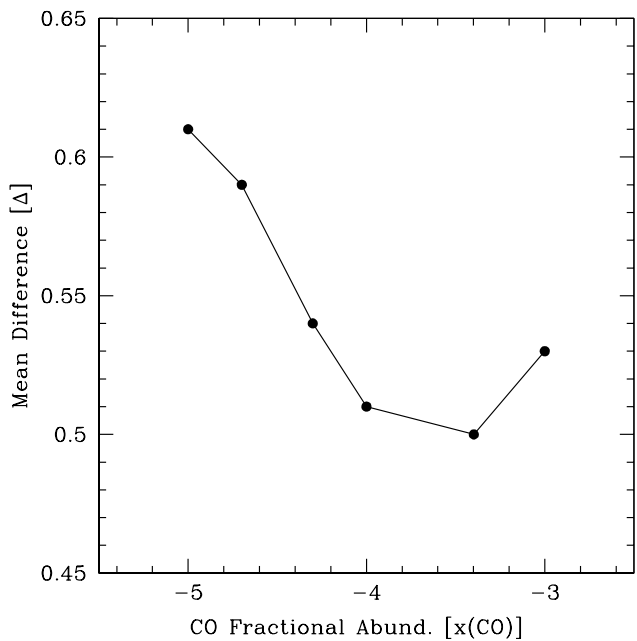}}
      \caption[]{Dependence of quality of fit as measured by the 
                 mean difference between predicted and observed line
		 strengths as a function of the CO fractional
		 abundance for $T>T_{\mathrm{CO}}=20$~K.
		 Notice the best fit for $x(\mathrm{CO}) \sim 10^{-4}$.
              }
         \label{figxco}
   \end{figure}
%
%

\subsection{Effects of CO abundance}

Previous authors (Sch\"oier et al. \cite{schoieretal2002};
Ceccarelli et al. \cite{ceccarellietal2000a}) have inferred (constant)
CO abundances in the range of $10^{-5} - 10^{-4}$.  As a result
the inner ($T>T_{\mathrm{des,CO}}$) CO abundance is varied,
keeping all other parameters the same as in our best fit model.
The results are presented in Fig. \ref{figxco}.

%
   \begin{figure}
      \resizebox{\hsize}{!}{\includegraphics{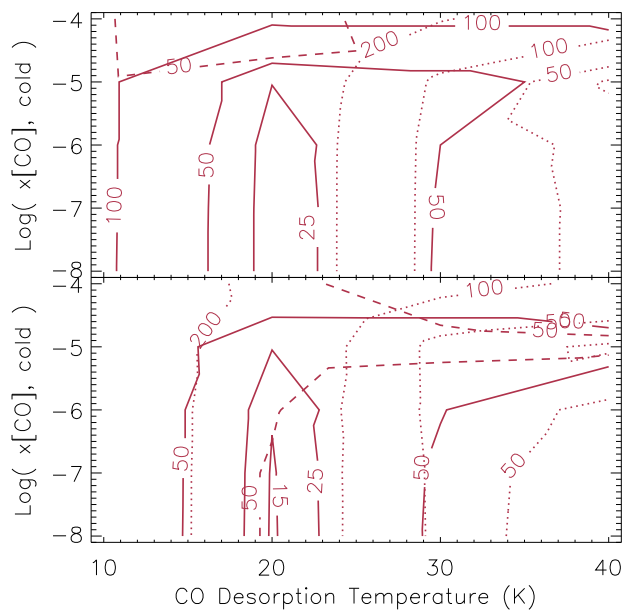}}
      \caption[]{Contours of the percentage difference between model
                 and observed CO line strengths, for various values of
		 CO desorption temperature ($T_{\mathrm{CO}}$), 
		 abundance of cold CO ($T<T_{\mathrm{CO}}$), and
		 abundance of warm CO ($T>T_{\mathrm{CO}}$) denoted
		 by the different line types.  The solid, dotted,
		 and dashed lines correspond to  
		 $x(\mathrm{CO,warm})=10^{-4}$, $5\times10^{-4}$, 
		 and $2\times10^{-5}$ respectively.
		 The top panel gives
		 the results for C$^{17}$O, and the bottom panel 
		 the average results for $^{13}$CO, C$^{17}$O, and
		 C$^{18}$O.  Notice that the C$^{17}$O results 
		 strongly favor 
		 $x(\mathrm{CO,warm})=10^{-4}$, as do the 
		 results averaged over the CO isotopomers.
              }
         \label{figco2panel}
   \end{figure}
%
%

%
   \begin{figure}
      \resizebox{\hsize}{!}{\includegraphics{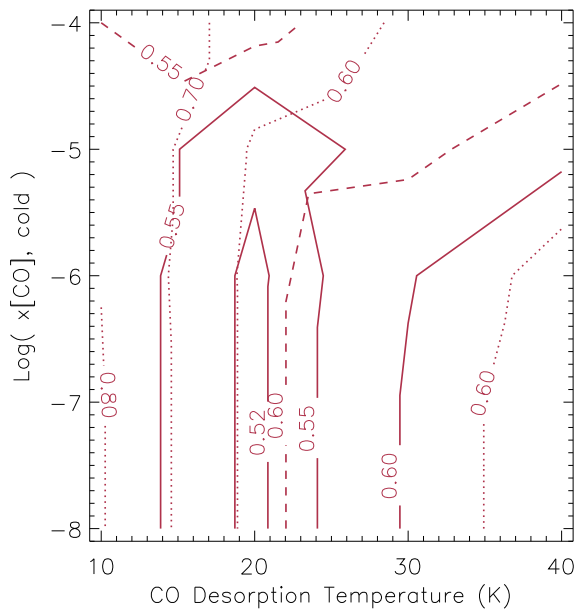}}
      \caption[]{Contours of quality of fit as measured by the 
                 mean difference between predicted and observed line
		 strengths over all species as a function of the 
		 CO desorption temperature ($T_{\mathrm{CO}}$), 
		 abundance of cold CO ($T<T_{\mathrm{CO}}$), and
		 abundance of warm CO ($T>T_{\mathrm{CO}}$) denoted
		 by the different line types.  The solid, dotted,
		 and dashed lines correspond to  
		 $x(\mathrm{CO,warm})=10^{-4}$, $5\times10^{-4}$, 
		 and $2\times10^{-5}$ respectively.
		 Notice the agreement with Fig. \ref{figco2panel}
                 in constraining $16<T_{\mathrm{CO}}<30$ K, 
		 $x(\mathrm{CO,cold})<10^{-5}$, and
		 $x(\mathrm{CO,warm}) \sim 10^{-4}$.
              }
         \label{figcosome}
   \end{figure}
%
%

As can be seen, CO abundances of $1-4 \times 10^{-4}$
are preferred.  Lower abundances produce too little
C$^{18}$O emission to be consistent with observations.  The best
fit -- based upon CO data alone -- is $x(\mathrm{CO}) = 10^{-4}$
where models match C$^{18}$O observations to within 2\%.
By $x(\mathrm{CO}) = 5 \times 10^{-5}$ and $4 \times 10^{-4}$, 
the discrepancy for the C$^{18}$O lines reaches 37\% and 95\%
respectively.  The effect on $\Delta$ is
smaller for two reasons:  first other species are included in
the mean difference, and second the comparison for HCO$^{+}$ and
CN are both improved as $x(\mathrm{CO})$ increases.

The HCO$^{+}$ and CN abundances both rely upon 
CO via ion-molecule reactions.
Since this relation is much more indirect, the roles of HCO$^{+}$ and
CN abundances in fixing the CO abundance are discounted, and 
$x(\mathrm{CO}) \sim 10^{-4}$ is preferred.  The constant abundance
of $\sim 4\times 10^{-5}$ inferred by Sch\"oier et al.\
(\cite{schoieretal2002}) in the empirical model can now be understood
as a weighted average of a very low CO abundance at $T<20$~K and a
higher abundance of $\sim 10^{-4}$ at $T>20$~K.

We have also considerd the combined effects of CO desorption
temperature, cold ($T<T_{\mathrm{CO}}$) CO abundance, and warm
($T>T_{\mathrm{CO}}$) CO abundance.  The results are presented in
Fig. \ref{figco2panel} where we plot the mean percentage difference
between model and observed line strengths for C$^{17}$O (top panel),
and all CO isotopomers (bottom panel).  Warm CO abundances are denoted
by the different line types, corresponding to $x(\mathrm{CO,warm}) =
10^{-4}, 5\times10^{-4}, 2\times10^{-5}$ respectively.  As can be
seen, the CO data strongly prefer $x(\mathrm{CO,warm}) \sim 10^{-4}$,
consistent with our results above.  This is confirmed
by the results in Fig. \ref{figcosome} where we plot contours of the
mean difference, $\Delta$, over all observed lines.  Again, models
with $x(\mathrm{CO,warm}) \sim 10^{-4}$ are preferred.

Perhaps even more interestingly, the results for both the CO and
general chemical network allow us to simultaneously constrain the
desorption temperature and cold CO abundance.  Taking uncertainties in
the observational data of $\pm 30$\% suggests $18 < T_{\mathrm{CO}}
\mathrm{(K)} < 23$ based upon the 25\% contour level in
Fig. \ref{figco2panel}.  Even if the uncertainties are larger, the
results in Figs. \ref{figco2panel} and \ref{figcosome} provide outer
limits of $16 < T_{\mathrm{CO}} \mathrm{(K)} < 30$.  There may
be a potential region of degeneracy in Figs. \ref{figco2panel} \&
\ref{figcosome} as the cold CO abundance appears to increase with 
increasing $T_{\mathrm{CO}}$.  It should be noted that the cold
CO abundance is still $\sim 10^{-5}$ in this case, a result that may
be explained by significant evaporation near $T_{\mathrm{CO}}$ and
some partial/gradual evaporation of CO presumably in an H$_{2}$O 
matrix at higher temperatures.  These results are
consistent with both the cut along $T_{\mathrm{CO}}$ and the
laboratory results discussed above.

Finally, there does seem to be evidence of depletion in the cold gas
for $T<T_{\mathrm{CO}}$.  The comparison for both the CO and overall
set of observed species suggest a relatively firm upper limit of
$x(\mathrm{CO,cold}) < 10^{-5}$.  The regions of best fit appear to
encompass values of 3-30 times less ($3 \times 10^{-7}$ to $3 \times
10^{-6}$).  The upper value of $10^{-5}$ signifies a depletion of
90\%, while the lower values correspond to 97\% and 99\% depletion
respectively.  While the exact level of depletion is uncertain, these
results do confirm a significant sink of gas-phase CO -- presumably as
ices onto dust grains in the cool exterior.  Such high levels of CO
depletion are consistent with those found in cold pre-stellar cores
(e.g., Bacmann et al.\ \cite{bacmannetal2003}) and the large abundances
of deuterated molecules detected in the outer envelope of IRAS
16293-2422 (van Dishoeck et al.\ \cite{vandishoecketal1995}; Loinard
et al.\ \cite{loinardetal2000}; Parise et al. \cite{pariseetal2002})

\subsection{Effects of H$_{2}$CO and CH$_{3}$OH depletion temperature}

   \begin{figure}
      \resizebox{\hsize}{!}{\includegraphics{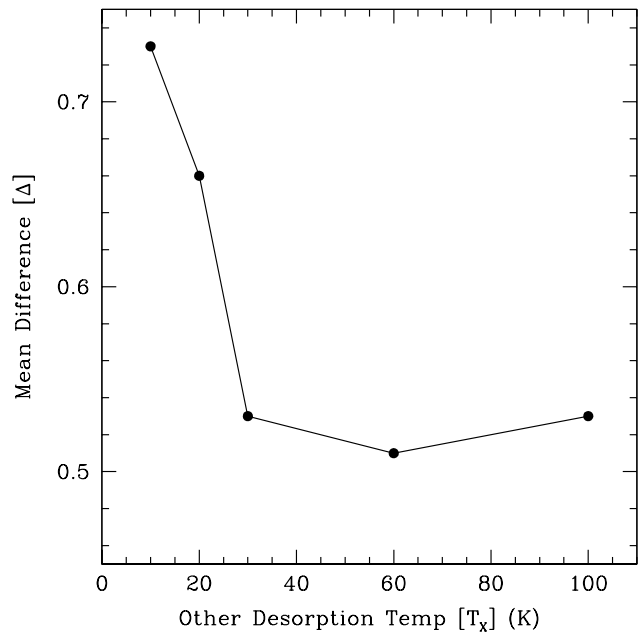}}
      \caption[]{Dependence of quality of fit as measured by the 
                 mean difference between predicted and observed line
		 strengths as a function of the temperature of 
		 H$_{2}$CO and CH$_{3}$OH desorption ($T_{X}$).
		 Notice the best fit for $T_{X} > 30$ K.
              }
         \label{figtx}
   \end{figure}
%
%

We also consider the effects of modifying the H$_{2}$CO and
CH$_{3}$OH desorption temperatures, $T_{X}$.  In the case of 
H$_{2}$CO which may be produced at some level in the 
gas phase (Doty et al. \cite{dotyetal2002}), this 
corresponds to the temperature at which significant 
production occurs.  The results are shown in Fig. \ref{figtx},
where the mean difference is plotted as a function of $T_{X}$.

To within the uncertainties of the observations and sophistication of
the models, the results for $T_{X}=30-100$ K are
indistinguishable.  Consequently, the results in Fig. \ref{figtx}
suggest T$_{X} > 30$ K.  
In the case of CH$_{3}$OH, which is almost
certainly formed on the grain surface (e.g., Tielens \& Hagen
\cite{tielenshagen1982}, Blake et al. \cite{blakeetal1987}, Doty et
al. \cite{dotyetal2002}), this is probably an indication of
desorption.  Desorption in this temperature range is consistent with
current chemical understanding of solid ices.  Sandford \& Allamandola
(\cite{sandfordallamandola1993b}) measure a pure CH$_{3}$OH desorption
temperature (in space) of $70-80$ K.  This is in keeping with the fact
that CH$_{3}$OH should have a lower desorption temperature than water
due to its weaker hydrogen bonding.
For comparison, a Clausius-Clapeyron
calculation which reproduces the water evaporation temperature well
suggests an evaporation temperature for CH$_{3}$OH of $\sim 87$ K
(Alsindi et al. \cite{alsindietal2003}).  
Likewise, these results are consistent with the empirical modeling of 
Sch\"oier et al. (\cite{schoieretal2002}), who found 
$T_X(\mathrm{CH}_{3}\mathrm{OH}) > 50$ K.
As a result, it is encouraging that
laboratory work, theoretical calculations, empirical models, 
and the results of this
work are all in agreement with a CH$_{3}$OH desorption temperature of
$60 < T(\mathrm{K}) <100$.

In the case of H$_{2}$CO, the meaning of this temperature is less
clear.  The deviation between the observed and predicted H$_{2}$CO
line strengths caused by the overproduction of H$_{2}$CO in the
model do not significantly change as $T_{X}$ is increased -- from 49\% at
$T_{\mathrm{des}}=10$ K to 44\% at $T_{\mathrm{des}}=100$ K.  While
Doty et al.  (\cite{dotyetal2002}) suggested it was possible that
gas-phase reactions could play a role in the H$_{2}$CO chemistry, they
did not identify any that would cause a significant `jump' in
abundances in this temperature range.  
While it is difficult to directly constrain the H$_{2}$CO 
abundance, it is clear that the modeling here is insensitive to the
amount of cold H$_{2}$CO, and that while no jump is required, a jump
due to desorption is not ruled out -- consistent with 
$T_{X} > 40$K as found by  
Sch\"oier et al. (\cite{schoieretal2002}).

\subsection{Outliers}

As can be seen in Fig. \ref{bestfitcomparison}, the three species 
OCS, C$_{2}$H, and HCS$^{+}$ are outliers in our models. These
species yield line strengths which diverge from the observations by factors
of $\sim 5$, $100$, $30$, and $100$ respectively.  Some deviation due
to radiative transfer, geometrical, and line of sight effects
is to be expected.
However, the differences for these three
species are discrepant from the other 
species considered, implying that while our adopted physical/chemical
structure is reasonable, our knowledge of
the chemistry is lacking.

In the case of OCS, the chemistry is uncertain, and many of
the reaction rates are estimates without significant laboratory
study.  As a result, it is not suprising that a 
discrepancy exists.  For C$_{2}$H, 
UV photodissociation in the outer region  -- while included -- 
is not significant
as most of the C$_{2}$H is produced above 100K.
The dominant production
is via recombination of C$_{2}$H$_{5}^{+}$, and reaction of
C$_{3}$H$_{2}^{+}$ with O.  Destruction is mainly through
the neutral-neutral reaction with O.
While a lower oxygen abundance can increase the C$_{2}$H abundance,
it also decreases the abundances of the other important
oxygen-bearing species such as SO, SO$_{2}$, and CH$_{3}$OH
to below the observations.  
On the other hand, 
it is interesting to note that the destruction
reaction with O is assumed to be temperature independent (MFW).
The existence of a reaction barrier or a temperature
dependent rate of collisions would both tend to decrease the 
destruction, which would have the effect of raising the C$_{2}$H
abundance closer to the observed levels.

The third outlier, HCS$^{+}$ follows an ionization balance
with CS via dissociative recombination, and
reactions with HCO$^{+}$, H$_{3}^{+}$, and H$_{3}$O$^{+}$.
However, raising $\zeta$ to $5 \times 10^{-15}$ s$^{-1}$ only changes 
the HCS$^{+}$ discrepancy by $\sim$ 3\%.  Such high values of the cosmic
ray rate raise the most sensitive ion -- HCO$^{+}$ -- 
to levels far above that observed.  While the major destruction
mechanism of dissociative recombination has a somewhat strong
($T^{-0.75}$) temperature dependence, this rate has been measured in the 
laboratory, and is considered to be accurate to within 25\% by MFW.
This leaves the production reactions of (HCO$^{+}$, H$_{3}^{+}$, and
H$_{3}$O$^{+}$) + CS as potential sources of uncertainty.  Each of these
rates are estimated.  As such, it would be useful to measure them in 
the laboratory to confirm the rates adopted by MFW.

Finally, 
while HNC only differs by a factor of 5 and is thus
not a major outlier,
the production of HNC is still not
fully understood (e.g.
Rodgers \& Charnley \cite{rodgerscharnley2001};
Charnley, Rodgers \& Ehrenfreund \cite{charnleyrodgersehrenfreund2001};
Liszt \& Lucas \cite{lisztlucas2001}).  
In our model, HNC is produced primarily through the dissociative
recombination of HCNH$^{+}$ and H$_{2}$NC$^{+}$.  In the recombination of
HCNH$^{+}$, where the branching fractions are taken to be 50\% for CN,
25\% for HCN, and 25\% for HNC, the HNCH$^{+}$
abundance is determined mostly by
an ionization equilibrium in which the primary production
paths are proton transfer between HNC and (HCO$^{+}$, H$_{3}$O$^{+}$,
H$_{3}^{+}$).  These are also the dominant destruction paths for HNC.
On the other hand, H$_{2}$NC$^{+}$ is formed by 
C$^{+} +$ NH$_{3} \rightarrow$ H$_{2}$NC$^{+}$, and
dissociatively recombines to form HNC and CN in a 10:1 ratio.
We note that while too much HNC is produced in our model, the CN
abundance is somewhat low.  This combination suggests that the
adopted branching fractions for the dissociative recombination
should perhaps be reinvestigated.  Adopting the 
results of Talbi \& Herbst (\cite{talbiherbst1998})
for C$^{+} + \mathrm{NH}_{3} \rightarrow \mathrm{HCNH}^{+}$ 
has little effect, suggesting further concentration on H$_{2}$NC$^{+}$
Alternatively, some HNC destruction route may be missing in the networks.

\subsection{Comparison with empirical model}

   \begin{figure*}
      \resizebox{\hsize}{!}{\includegraphics{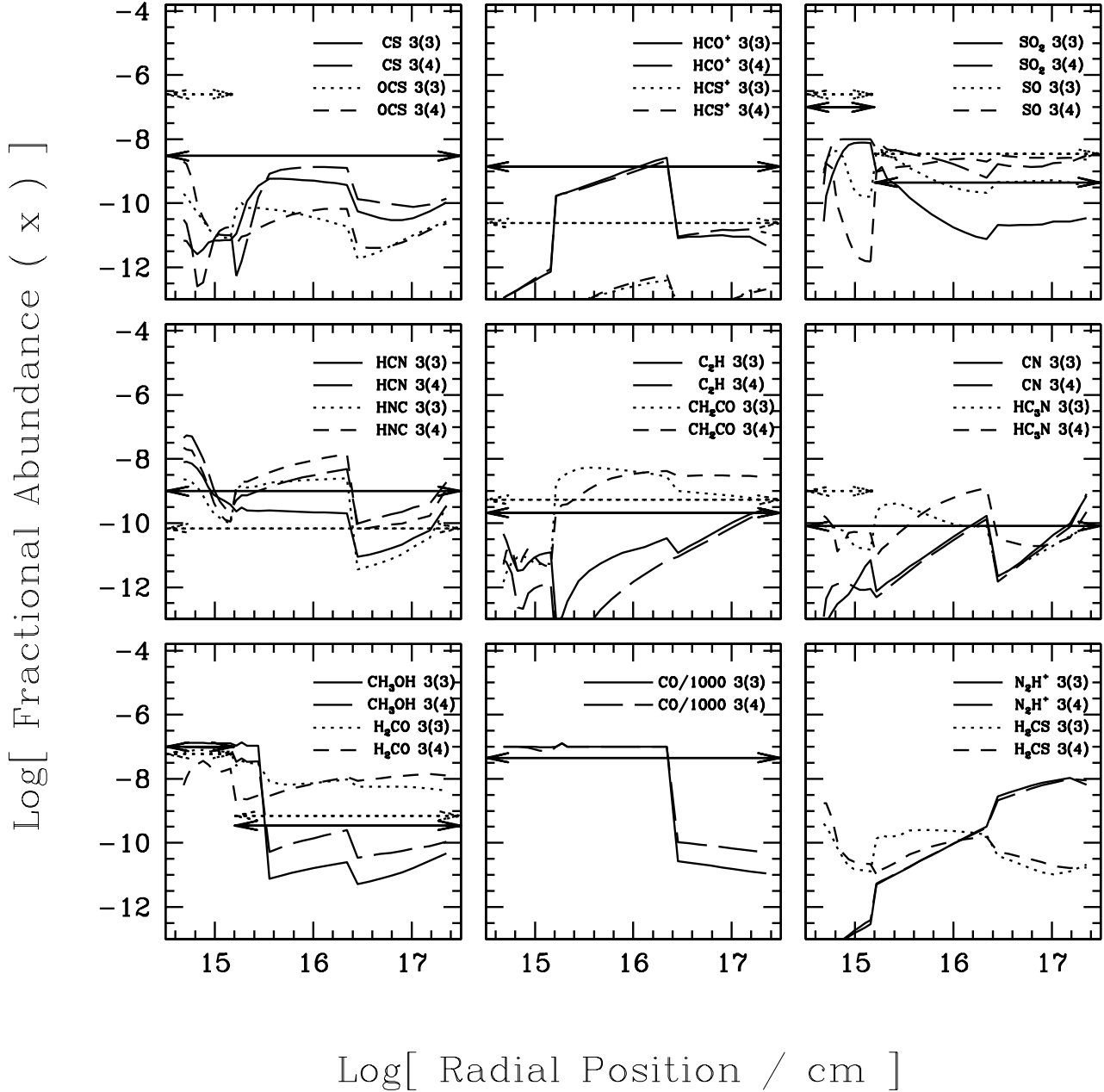}}
      \caption[]{Radial abundance profiles for
                 the species considered in the text 
		 for $t=3 \times 10^{3}$ and $3 \times 10^{4}$ years.
		 The results of the empirical model of 
		 Sch\"oier et al. (\cite{schoieretal2002}) are
		 given by the lines with arrows, for comparison.
              }
         \label{plot.8panel}
   \end{figure*}
%
%

In principle, the empirical modeling approach should -- with
sufficient parameter variation -- mimic the results of 
the full chemical modeling.  In practice, it is difficult
to parametrically vary all species in a sufficiently meaningful
yet complete manner.  As discussed in \S 1, the approach adopted
by many authors is to treat abundances as either
constant, or as piecewise constant with `jumps' at
appropriate temperatures.  This was the approach taken 
in Sch\"oier et al (\cite{schoieretal2002}).  It is interesting
to compare the inferred abundance distributions from the
empirical modeling from the more detailed results of the
full chemical modeling.

Quantitatively, the Sch\"oier et al. (\cite{schoieretal2002})
empirical abundances reproduce the observations with
$\Delta=0.24$, about half of our best
fit model, $\Delta=0.51$.  This is, however, not a suprise.   In the
Sch\"oier et al. (\cite{schoieretal2002}) model, there are many 
more free parameters as the abundances for each of the observed
species are varied both above and below the assumed desorption
temperature.  Furthermore, these variations are done without
respect to constraints on the chemical network or evolution time.
On the other hand, we directly specify the initial abundances for
only three of the observed species (CO, H$_{2}$CO, and CH$_{3}$OH), 
and are constrained by the chemical network and its evolution.  
Furthermore, as discussed previously, the majority of the inputs
to the chemical network are taken directly from agreement reached
on a high-mass hot-core source, AFGL 2591.
Taken together, 
these results are strongly encouraging as the
chemical network comparison gives a good fit for significantly
fewer direct parameter variations, confirms the proposed age of
the source from physical evolution models, and directly tests the
validity and extensibility of the chemical networks. 

As a more direct comparion,
the radial abundance profiles for the species
discussed in this paper are presented in 
Fig. \ref{plot.8panel}, where
the two limiting times of $3 \times 10^{3}$ years and 
$3 \times 10^{4}$ years are plotted.
In general, the abundances inferred from the empirical
modeling are grossly consistent with those from the more
detailed full chemical modeling.  For most species with 
constant abundances, the inferred abundances in mostly the 
outer envelope are equivalent to within a factor of 3--10 at our best-fit
time (here taken to be that intermediate
between the two limits). In some cases, e.g. CN, CS, and CH$_{2}$CO, the 
chemical model abundance
oscillates with position in the cloud and the empirical value is simply 
a rough average of these complicated profiles.
The most significant discrepancies are for OCS, HCS$^{+}$,C$_{2}$H, 
and HNC, which are discussed above.

Even more interesting is the comparison of the jump models.  In
general, those species which show significant spatial variation in the
full chemical model are represented as `jumps' in the Sch\"oier et
al. (\cite{schoieretal2002}) empirical model.  Of these, the empirical
and full chemical models generally agree to within a factor of 3 or
so.  As discussed in \S 3.4, the inferred CO abundance can be
understood in such a `jump' model.  The significant discrepancy is
H$_{2}$CO, which the full chemical model predicts to have only a small
jump, while the empirical model infers a cold, outer abundance some
15 times lower.

\subsection{Predictions for future observations}

   \begin{figure*}
      \resizebox{\hsize}{!}{\includegraphics{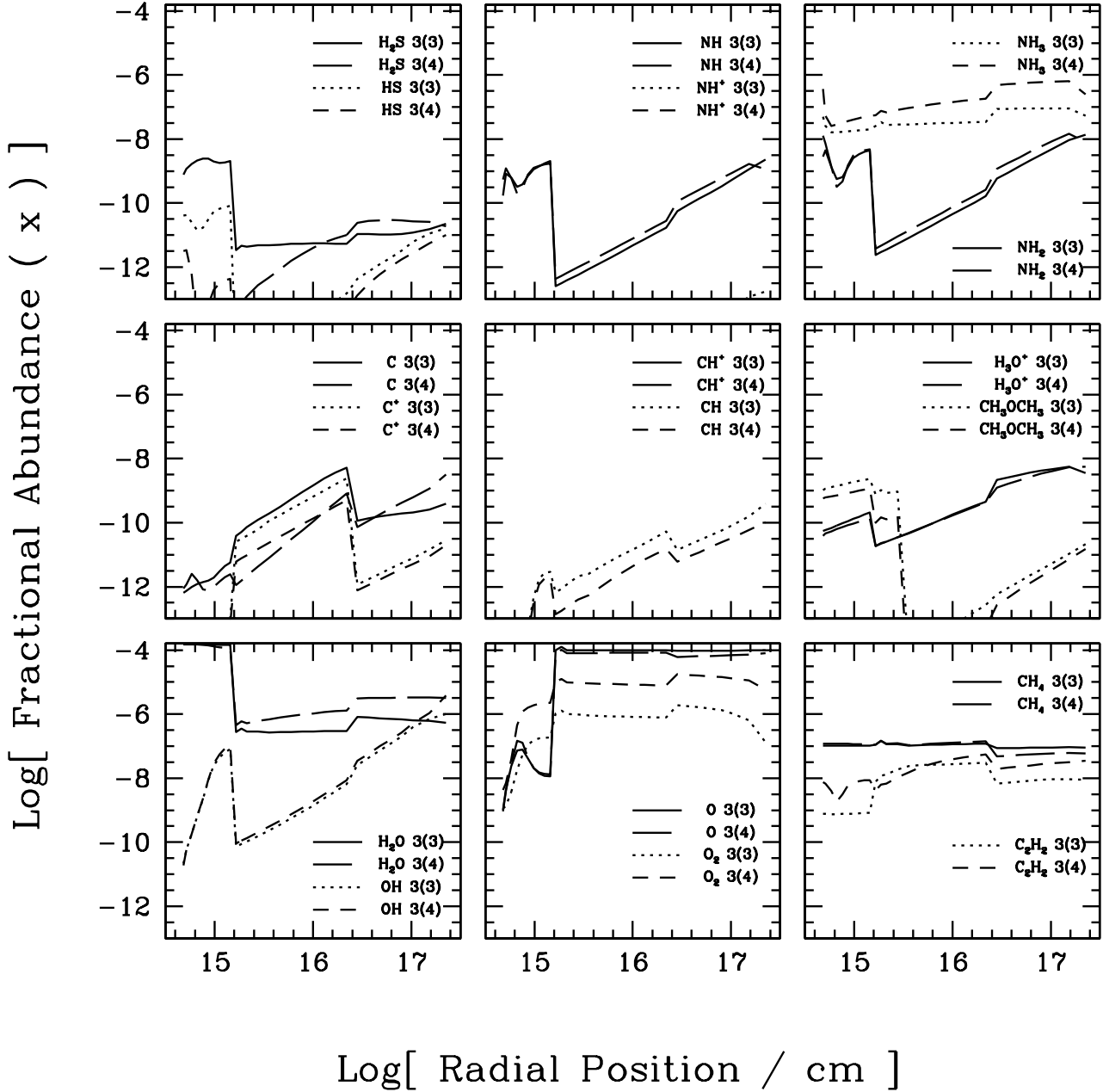}}
      \caption[]{Radial abundance profiles for
                 the some of the more abundant species 
		 that may be targets for future observations.
		 The data are reproduced for 
		 $t=3 \times 10^{3}$ and $3 \times 10^{4}$ years.
              }
         \label{plot.9panel}
   \end{figure*}
%
%

While the model presented is able to simultaneously match 
the SED and many of the molecular line observations, a 
significant test will be the predictions it makes that
can be studied by future facilities.  In particular, 
with CARMA, SIRTF, and SOFIA data to be available in the
next few years, and the upcoming leaps in resolution and
sensitivity from Herschel and ALMA, it will be possible to 
probe many of the transitions and much of the spatial 
structure of IRAS 16293.

To aid such future observations, the radial distributions
of a number of interesting species are shown in Fig. \ref{plot.9panel}.
Also, column densities predicted by the
model toward IRAS 16293-2422  for $t=3 \times 10^{3}$ years and
$3 \times 10^{4}$ years are given in Table \ref{columndensities}.  
This includes, in particular,
radial column densities given by 
$N_{X,\mathrm{rad}} = \int n_{X}(r) dr$, where $n_{X}(r)$ is 
the density of species $X$ as a function of $r$.  We also 
include the column density averaged over a 15 arcsecond
beam toward IRAS 16293, given by
$N_{X,\mathrm{beam}} = \int \int n_{X}(z,b) dz G(b) 2\pi b db / \int G(b)
2 \pi b db$, where $b$ is the impact parameter, and $G(b)$ is the beam
response function.  The column densities are sorted from highest
to lowest in the 15 arcsecond beam at $t=3 \times 10^{3}$ years,
and continue down to $\sim 10^{13}$ cm$^{-2}$.

   \begin{table}
      \caption[]{IRAS 16293-2422 predicted column densities (cm$^{-2}$)} 
         \label{columndensities}
         \begin{tabular}{lllll}
            \hline
   {time(yrs)}   &  $3E3$  & $3E3$  & $3E4$  & $3E4$  \\
            Species & N$_{\mathrm{rad}}$ & N$_{\mathrm{beam}}$
                    & N$_{\mathrm{rad}}$ & N$_{\mathrm{beam}}$ \\
            \hline
H$_2$         &     2.0E+24   &     2.0E+23   &     2.0E+24   &     2.0E+23\\
He         &     3.4E+23   &     3.3E+22   &     3.4E+23   &     3.3E+22\\
H          &     2.2E+19   &     2.0E+19   &     6.0E+19   &     6.1E+19\\
O          &     5.7E+19   &     1.4E+19   &     1.0E+19   &     1.8E+18\\
N$_2$         &     1.4E+20   &     1.4E+19   &     1.4E+20   &     1.3E+19\\
CO         &     1.8E+20   &     1.0E+19   &     1.6E+20   &     1.0E+19\\
O$_2$         &     8.1E+18   &     2.2E+18   &     4.5E+19   &     8.5E+18\\
H$_2$O        &     1.8E+20   &     8.2E+17   &     1.2E+20   &     2.6E+17\\
CO$_2$       &     4.7E+19   &     1.5E+17   &     7.4E+19   &     3.0E+17\\
N          &     3.4E+17   &     1.3E+17   &     2.8E+17   &     4.8E+16\\
NO         &     1.2E+17   &     9.1E+16   &     3.5E+17   &     3.4E+17\\
NH$_3$        &     2.0E+17   &     6.5E+16   &     2.9E+17   &     7.8E+16\\
OH         &     5.2E+16   &     3.0E+16   &     1.0E+17   &     9.7E+16\\
CH$_4$        &     2.3E+17   &     1.9E+16   &     2.3E+17   &     1.1E+16\\
C$_2$H$_2$       &     2.2E+16   &     6.4E+15   &     2.4E+16   &     4.5E+15\\
e$^{-}$          &     2.1E+16   &     5.2E+15   &     2.5E+16   &     5.7E+15\\
C$_2$H$_4$       &     5.2E+16   &     3.8E+15   &     9.7E+14   &     7.9E+12\\
Fe$^{+}$        &     1.4E+16   &     3.1E+15   &     1.7E+16   &     3.5E+15\\
H$_2$CO       &     2.9E+16   &     1.8E+15   &     1.7E+15   &     3.9E+14\\
HNC        &     1.6E+16   &     1.0E+15   &     2.6E+16   &     3.3E+14\\
Fe         &     2.6E+16   &     7.9E+14   &     2.3E+16   &     4.1E+14\\
C$_2$H$_6$       &     2.3E+15   &     7.0E+14   &     3.3E+12   &     1.7E+12\\
CHOOH      &     6.1E+15   &     6.5E+14   &     1.2E+16   &     7.9E+14\\
CH$_2$CO      &     1.2E+15   &     6.1E+14   &     4.2E+13   &     2.7E+13\\
CH$_3$OH      &     1.3E+17   &     5.2E+14   &     1.8E+16   &     3.9E+13\\
HNO        &     5.2E+14   &     5.1E+14   &     1.1E+16   &     1.2E+16\\
H$_3^+$        &     4.7E+14   &     4.8E+14   &     6.5E+14   &     6.7E+14\\
SO         &     2.0E+15   &     4.7E+14   &     1.3E+15   &     2.0E+14\\
N$_2$H$^+$       &     4.2E+14   &     4.4E+14   &     5.6E+14   &     5.9E+14\\
NH$_2$        &     5.7E+15   &     4.3E+14   &     1.5E+16   &     9.1E+15\\
HCN        &     3.4E+16   &     4.0E+14   &     5.5E+16   &     1.7E+14\\
SO$_2$        &     1.1E+16   &     3.5E+14   &     1.4E+16   &     8.9E+14\\
H$_3$O$^+$       &     3.6E+14   &     2.6E+14   &     1.2E+14   &     5.6E+13\\
S          &     1.6E+15   &     2.1E+14   &     2.6E+14   &     1.5E+13\\
CS         &     4.2E+14   &     1.3E+14   &     2.0E+14   &     9.6E+13\\
HCO$^+$       &     3.2E+14   &     1.2E+14   &     4.5E+14   &     1.6E+14\\
C$_3$H        &     1.5E+14   &     1.2E+14   &     1.7E+13   &     7.1E+12\\
HC$_3$N       &     2.6E+14   &     7.8E+13   &     5.1E+13   &     9.3E+12\\
He$^+$       &     6.5E+13   &     6.7E+13   &     5.2E+13   &     5.3E+13\\
C          &     7.8E+13   &     6.7E+13   &     6.3E+12   &     4.4E+12\\
CH$_3$CHO     &     1.6E+14   &     6.6E+13   &     1.7E+13   &     6.4E+12\\
NH         &     1.2E+15   &     4.8E+13   &     1.7E+15   &     6.9E+14\\
C$_3$H$_2$       &     7.8E+13   &     4.7E+13   &     4.9E+13   &     3.0E+13\\
N$^+$         &     4.5E+13   &     4.6E+13   &     2.4E+13   &     2.5E+13\\
C$_4$H        &     6.5E+13   &     4.3E+13   &     4.0E+13   &     2.5E+13\\
H$^+$         &     4.4E+13   &     4.1E+13   &     1.9E+13   &     1.9E+13\\
C$_3$H$_3$       &     1.0E+14   &     3.3E+13   &     7.3E+12   &     1.0E+12\\
C$_2$H$_5$OH     &     2.4E+13   &     2.4E+13   &     3.9E+10   &     1.1E+10\\
C$^+$         &     4.3E+13   &     2.3E+13   &     1.1E+13   &     6.4E+12\\
O$_2^+$        &     2.1E+13   &     1.7E+13   &     3.0E+13   &     2.5E+13\\
NO$_2$       &     7.5E+13   &     1.6E+13   &     2.7E+14   &     2.7E+14\\
H$_2$CS       &     7.9E+14   &     1.6E+13   &     5.1E+14   &     7.6E+11\\
NH$_4^+$       &     1.8E+13   &     1.5E+13   &     1.8E+13   &     1.4E+13\\
C$_2$S        &     3.4E+13   &     1.5E+13   &     1.8E+12   &     1.1E+12\\
HNO$^+$       &     8.8E+12   &     9.2E+12   &     4.4E+13   &     4.7E+13\\
CN         &     1.4E+13   &     9.2E+12   &     1.7E+12   &     9.0E+11\\
            \hline
     \end{tabular}\\
   \end{table}

Finally, it is intruiging to speculate on the effects and potential
observability of a collapsing cloud.
The evolution of a gas parcel in a protostellar envelope
passing from a low temperature / density
to a high temperature / density through infall has been studied
by CHT96, and Rodgers \& Charnley (\cite{rodgerscharnley2003}) 
among others.
In the outer regions, cool ion-molecule chemistry dominates.
As the gas heats while infalling, each adsorbed species passes
through a sublimation front.
In the interior warm neutral chemistry
can take place, 
for example the conversion of gas-phase oxygen
to H$_{2}$O 
(e.g., CHT96; Charnley \cite{C97}), 
and the production of significant complex cyanogens and hydrocarbons
(e.g., Rodgers \& Charnley \cite{rodgerscharnley2001}; 
Doty et al. \cite{dotyetal2002}). 
In free-fall collapse of low-mass YSOs, the dynamical timescale in the warm
interior is less than the chemical timescale, leading to 
inner core abundances that mirror those in the cool exterior.
The observations of a wide variety of complex daughter species
observed in these warm interiors 
(e.g. Cazaux et al. \cite{cazauxetal2003})
can only be understood if the 
collapse/infall is ``slowed'' to timescales over at least
$10^{4}$ years 
(Rodgers \& Charnley \cite{rodgerscharnley2003})
-- a scenario more in-line with our adopted
static case.  Since the chemistry encodes the
temperature/density temporal evolution of the gas, small-scale
spatial variations in abundances should, in principle, be able
to distinguish between various dynamical scenarios such as 
static, Shu (\cite{shu1977}) collapse, 
Larson-Penston (Larson \cite{larson1969}; Penston \cite{penston1969})
infall, etc.  
This will, however, require a next generation of modeling
that includes detailed dynamics, thermal balance, chemistry, \&
radiative transfer, as well as    
high spatial
and spectral resolution observations with instruments 
such as Herschel, ALMA, CARMA, and SOFIA to probe multiple
lines at $\sim 100$ AU resolution.

\section{Conclusions}
We have constructed detailed thermal and gas-phase
chemical models for IRAS 16293-2422 based upon the physical model
of Sch\"oier et al. (\cite{schoieretal2002}). 
These models were used to probe the validity of the proposed
physical structure, as well as study the chemical evolution of the source,
and to test the application of our combined `hot-core'/envelope chemistry
model of AFGL 2591 to a low-mass `hot-core'-like source.  
In particular, we find that:
\begin{enumerate}

\item The combined application of a physical, thermal, and
chemical model with detailed radiative transfer is a powerful
tool in constraining the structure and evolution of 
depth-dependent sources.

\item The time and position dependent model of Doty et al. 
(\cite{dotyetal2002}) can be meaningfully applied to 
IRAS 16293-2422, yielding results qualitatively similar to 
the massive YSO AFGL 2591.

\item The best fit for IRAS 16293-2422 occurs for
times in the range $3\times10^{3} < t(\mathrm{yrs}) < 3 \times 10^{4}$,
consistent with existing infall models (Sect. 3.1).

\item The best-fit ionization rate 
in IRAS 16293-2422 is high, $\sim 5 \times 10^{-16} - 10^{-15}$ s$^{-1}$,
compared with previous results for dense clouds.
We propose that this may be due to X-ray emission from the central sources
(Sect. 3.2).

\item Our best fit suggests that important CO desorption occurs
at low temperatures, $\sim 20$ K, and constrained to the range 
$15 < T_{\mathrm{CO}} \mathrm{(K)} < 40$.   Some solid CO may 
remain at higher temperatures for these timescales.
These results are in agreement with recent laboratory data of
CO on -- but not mixed with -- a water ice  (Sect. 3.3).

\item We can also constrain the warm ($T>T_{\mathrm{CO}}$) and cold
($T<T_{\mathrm{CO}}$) CO abundances to $x(\mathrm{CO,warm}) \sim
10^{-4}$ and $x(\mathrm{CO,cold}) < 10^{-5}$ and most probably $<
10^{-6}$.  These results are reflected in both the CO lines and in the
results from the greater chemical network, and suggest significant
($>$ 90\%) depletions at low temperatures (Sect. 3.4).

\item CH$_{3}$OH appears to desorb at temperatures
$60 < T(\mathrm{K}) < 100$, consistent with laboratory
expectations.  On the other hand, the comparison between the
H$_{2}$CO predictions and observations are 
insensitive to the amount of cold H$_{2}$CO present (Sect. 3.5).

\item The chemistry of HNC, C$_{2}$H, and HCS$^{+}$ may not
be fully understood.  In particular, it would be useful to 
measure the branching fraction of dissociative recombination
of H$_{2}$NC$^{+}$, the temperature dependence of the reaction
O$+$C$_{2}$H, and the ion-molecule reaction rates 
(HCO$^{+}$, H$_{3}^{+}$, H$_{3}$O$^{+}$) + CS (Sect. 3.6).

\end{enumerate}

\begin{acknowledgements}
      The authors are grateful to Jes J{\o}rgensen and Helen Fraser
      for fruitful discussions.  We thank the referee for 
      comments which improved the manuscript.
      This work was partially supported
      under grants from The Research Corporation (SDD), and the
      Netherlands Organisation for Scientific Research (NWO) through
      grant 614.041.004.  FLS further acknowledges financial support
      from the Swedish Reseach Council.  
      Astrochemistry at Leiden is supported
      through an NWO Spinoza award.
\end{acknowledgements}

\end{document}